\documentclass[9pt,conference]{IEEEtran}

\usepackage[preprint]{waspaa25}

\usepackage{bm} 
\usepackage{graphicx} 
\usepackage[T1, OT1]{fontenc}
\DeclareTextSymbolDefault{\dh}{T1}
\usepackage{hyperref}

\newcommand\blfootnote[1]{%
  \begingroup
  \renewcommand\thefootnote{}\footnote{#1}%
  \addtocounter{footnote}{-1}%
  \endgroup
}

\usepackage{pifont}
\newcommand{\cmark}{\ding{51}}%
\newcommand{\xmark}{\ding{55}}%

\title{MB-RIRs: a Synthetic Room Impulse Response Dataset with Frequency-Dependent Absorption Coefficients}

\name{Enric Gus\'o,$^{1,2}$
      Joanna Luberadzka,$^{2}$
      Umut Sayin,$^{2}$
      Xavier Serra$^{1}$}
 \address{$^1$ Universitat Pompeu Fabra, Music Technology Group, Barcelona \\ enric.guso@upf.edu, xavier.serra@upf.edu\\
          $^2$ Eurecat, Centre Tecnològic de Catalunya, Tecnologies Multimèdia, Barcelona \\
           joanna.luberadzka@eurecat.org, umut.sayin@eurecat.org\\
 }

\begin{document}

\maketitle

\begin{abstract}
    We investigate the effects of four strategies for improving the ecological validity of synthetic room impulse response (RIR) datasets for monoaural Speech Enhancement (SE). We implement three features on top of the traditional image source method-based (ISM) shoebox RIRs: multiband absorption coefficients, source directivity and receiver directivity. We additionally consider mesh-based RIRs from the SoundSpaces dataset. We then train a DeepFilternet3 model for each RIR dataset and evaluate the performance on a test set of real RIRs both objectively and subjectively. We find that RIRs which use frequency-dependent acoustic absorption coefficients (MB-RIRs) can obtain +0.51dB of SDR and a +8.9 MUSHRA score when evaluated on real RIRs. The MB-RIRs dataset is publicly available for free download.
\end{abstract}

\section{Introduction}
\label{sec:intro}
Speech enhancement (SE) is the combination of processes like dereverberation, denoising, declipping and bandwidth extension. In the last decade, deep learning-based SE methods have shown to be very effective, a success partially driven by the use of bigger and more diverse training datasets ---e.g. the ones used in the Deep Noise Suppression Challenges (DNS) \cite{dns5}. This data scaling has produced models that generalize better and mitigate cross-dataset performance differences like the ones in \cite{richter2023speech, kadiouglu2020empirical}. On top of scaling size and variety, speech and noise datasets have also been extended to a sampling rate of 48kHz, further improving the results \cite{schroter2023deepfilternet3, zhang2023toward, zhang2024urgent}. Generally speaking, the traditional approach for augmenting data is to convolve the speech signals with Room Impulse Responses (RIRs), simulating sounds in different acoustic environments. 

However, most RIRs currently used in the state-of-the-art are still the ones from the early DNS challenges \cite{ko2017study} ---i.e. shoebox-like rooms with a single acoustic absorption coefficient for the entire frequency spectrum. They are typically rendered through the image source method and at 16kHz sampling rate, and were originally intended to train automatic speech recognition systems. In fact, increasing the realism of the RIRs has been shown to improve speech recognition for augmented reality glasses \cite{arakawa2024quantifying} or keyword spotting \cite{bezzam2020study} even as a post-processing augmentation step \cite{ratnarajah2021ts}, so the same could happen for the SE task.

More recently, the URGENT Challenge \cite{zhang2024urgent} acknowledges the importance of RIR generalization by using real recorded RIRs for evaluation. While it constitutes an important step for improving the ecological validity of SE models, it still uses DNS5 RIRs for training, so their models might be confined to that particular RIR coverage. The topic is also covered in \cite{GenDA2025_RoomAcoustics}, with a focus on augmenting existing RIRs with generative models.

\blfootnote{  \hrule width 0.4\linewidth 
  \vspace{0.6em} 
The research leading to these results has received funding from the European union's Horizon Europe programme under grant agreement No 101017884 - GuestXR project. 

Dataset publicly available at \href{https://doi.org/10.5281/zenodo.15773093}{\textit{https://doi.org/10.5281/zenodo.15773093}}}

Beyond the monoaural SE task, mesh-based RIR data generation and refinement methods have been proposed in \cite{tang2022gwa, kelley2024rir}, for tasks such as audio-visual navigation. An example of this is the SoundSpaces dataset \cite{chen2022soundspaces}, which contains a set of RIRs from rooms with complex geometries, providing a grid of positions for each room, and is rendered by path-based methods on 3D meshes.  However, the performance of SE models trained on these RIRs has not been evaluated yet.

In this work we focus on the effects of RIRs used in training for SE. We evaluate five models trained on the same speech and noise datasets but on six different RIR datasets (two baselines, three of our own making and one from SoundSpaces), with the intention of answering whether models benefit from using RIRs with frequency dependent (multiband) absorption coefficients, whether modeling the receiver directivity or the source directivity helps to generalize in monoaural models \cite{arakawa2024quantifying, bezzam2020study}, and whether mesh-based modeling might be helpful or more suitable than Image Source Method (ISM) \cite{bezzam2020study}. To this end, we propose an evaluation of the following RIR datasets, which are summarized above in Table \ref{tab1}.

\begin{itemize}
    \item  \textbf{DNS5}: state of the art (SOTA) baseline from \cite{ko2017study}.
    \item \textbf{SB}: we re-create DNS5 at 48kHz sampling rate using \textbf{s}ingle-\textbf{b}and absorption coefficients on the shoebox room material.
    \item \textbf{MB}: we use \textbf{m}ulti-\textbf{b}and absorption coefficients instead.
    \item \textbf{REC+MB}: we add \textbf{rec}eiver directiviy by using Head Related Transfer Functions (HRTFs) at the rendering stage.
    \item \textbf{SRC+REC+MB}: we add \textbf{s}ou\textbf{rc}e directivity by modelling average human speech directivity in the ISM.
    \item \textbf{SSPA}: we use mesh-based RIRs from \textbf{S}ound\textbf{Spaces} instead.
\end{itemize}

\begin{table}[!tp]
\renewcommand{\arraystretch}{1.3} 
\caption{Summary of the evaluated synthetic RIR datasets. $f_{s}$ stands for sampling rate, $rec$ for receiver, $src$ for source, $render$ for rendering method and $T60$ for reverberation time.}

\begin{center}

\begin{tabular}{lll|ll|l}
                                &                          &                       & \multicolumn{2}{l|}{directivity} &        \\ \cline{2-6} 
\multicolumn{1}{l|}{}           & \multicolumn{1}{l|}{$f_{s}$}  & $T60$ & \multicolumn{1}{l|}{$rec$}  & $src$  & render \\ \hline
\multicolumn{1}{l|}{DNS5}       & \multicolumn{1}{l|}{16k} & single                & \multicolumn{1}{l|}{\xmark}   & \xmark   & ISM    \\ \hline
\multicolumn{1}{l|}{SB}         & \multicolumn{1}{l|}{48k} & single                & \multicolumn{1}{l|}{\xmark}   & \xmark   & ISM    \\ \hline
\multicolumn{1}{l|}{MB}         & \multicolumn{1}{l|}{48k} & multi                 & \multicolumn{1}{l|}{\xmark}   & \xmark   & ISM    \\ \hline
\multicolumn{1}{l|}{REC+MB}     & \multicolumn{1}{l|}{48k} & multi                 & \multicolumn{1}{l|}{\cmark}  & \xmark   & ISM    \\ \hline
\multicolumn{1}{l|}{SRC+REC+MB} & \multicolumn{1}{l|}{48k} & multi                 & \multicolumn{1}{l|}{\cmark}  & \cmark  & ISM    \\ \hline
\multicolumn{1}{l|}{SSPA}       & \multicolumn{1}{l|}{48k} & single                & \multicolumn{1}{l|}{\cmark}  & \xmark   & mesh  

\end{tabular}
\label{tab1}
\end{center}
\end{table}

\section{Training Datasets}
\label{sec:datasets}
Regarding speech and noise, we have used the whole DNS5 which contains emotional speech, readings from English audiobooks using different accents, singing from VocalSet, French, Spanish, Italian, Russian and German speech from M-AILABS Speech, German speech from Wikipedia, Spanish from SLR73, SLR61, SLR39, SLR75, SLR74 and SLR71 sets, which add up to a total of 583k speech utterances, and 63k noise samples from FreeSound and AudioSet \cite{dns5} wich add up to 1315 and 177 hours respectively. We have split them into 70\% for training, and the remaining 30\% split equally between validation and test sets, avoiding cross-set speaker contamination when possible. Below we describe the RIR training sets.

\subsection{Baseline RIRs (DNS5)}
\label{ssec:dns5}
We have taken as baseline 60k RIRs from DNS5 (synthetic RIRs from SLR26 and real RIRs from SLR28, originally at a 16kHz sampling rate and upsampled to 48kHz). We have distributed small, medium, and large rooms into three splits: 70\%, 15\% and 15\%  corresponding to training, validation and test sets.

\subsection{Single-band absorption coefficients RIRs (SB)}
\label{ssec:singleband}
With the intention of obtaining a fair comparison between the different strategies, we have created an approximation of the baseline DNS5 RIRs dataset using the Multichannel Acoustic Signal Processing (MASP) library \cite{masp, politis2016microphone}. Specifically, we have generated 60k shoebox-like RIRs with single-band absorption coefficients, this time rendering at 48kHz sampling rate. The geometric configurations of these RIRs are described in Table \ref{tab:rooms}: room dimensions $\boldsymbol{r}$ have been sampled from uniform distributions, with $r_{x}$ affecting $r_{y}$ in order to avoid corridor-like geometries. Single-band absorption coefficients for all six walls of the shoebox-like room are computed via Sabine's formula, using $\boldsymbol{r}$  and  reverberation time $T60$. In contrast to the multiband case (see \ref{ssec:multiband}), where $\boldsymbol{T60}$ is a vector defining reverberation time in each frequency band, in the single-band case $T60$ is a scalar computed as the mean of $\boldsymbol{T60}$ vector. $T60$ used for SB can be approximated by a normal distribution. Receiver coordinates $\boldsymbol{rec}$ have also been set randomly for every room, avoiding to place them too close to the walls. The sources $\boldsymbol{src}$ have been placed around the receivers at a distance between 0.5 and 3 meters and in front of them. We have used this same set of room configurations for the rest of the evaluated datasets except for the pre-computed SoundSpaces dataset SSPA.

\begin{table}[b]
\caption{Random room configurations for SB, MB, REC+MB, SRC+REC+MB. Angles are in degrees, vectors are in bold.}

\centering
\begin{tabular}{l|l}
\hline

                                             &                                                     \\
$r_{x}=\mathcal{U}(3 ,  30)$                 & $rec_x=\mathcal{U}(0.35r_{x}, 0.65r_{x})$          \\
$r_{y}=r_{x} \cdot \mathcal{U}(0.5 ,  1) $   & $rec_y=\mathcal{U}(0.35r_{y}, 0.65r_{y})$          \\
$r_{z}=\mathcal{U}(2.5,  5)   $              & $rec_z=\mathcal{U}(1, 2)$                          \\
                                             &                                                     \\
$||\boldsymbol{rec} - \boldsymbol{src}|| =\mathcal{U}(0.5, 3)$     & $rec_{\phi}=\mathcal{U}(-45, 45)$                \\
$\angle \boldsymbol{rec}, \boldsymbol{src} = \mathcal{U}(-45, 45)$ & $rec_{\theta}=\mathcal{U}(-10, 10)$                  \\
                                             &                                                     \\

$T60 \approx \mathcal{N}(0.4, 0.014) $ & $T60= \frac{1}{N}\sum_{n}\boldsymbol{T60}$ \\
$\boldsymbol{T60} = {Gamma}(\boldsymbol{\alpha}, \boldsymbol{\beta})$ &             \\ 

                                             &                                               \\ \hline
\end{tabular}
\label{tab:rooms}
\end{table}

\subsection{Multiband absorption coefficients RIRs (MB)}
\label{ssec:multiband}

\begin{figure}[t]
    \centering
    \centerline{\includegraphics[trim=25 11 20 8, width=\columnwidth, clip]{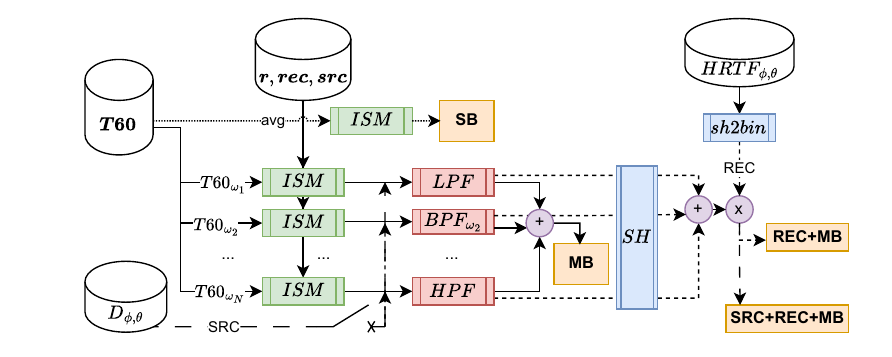}}
    \caption{RIR generation setup: the same set of geometric configurations is shared among the different datasets. Continuous line follows the MB pipeline, dotted line is for SB, short dashed line stands for the MB+REC modeling and long dashed line stands for the SRC modeling.}
    \label{fig:pipeline} 
\end{figure}

The first of the three strategies we have proposed is to increase the dataset coverage by using multiband absorption coefficients $\boldsymbol{T60}$ \cite{bezzam2020study} instead of a single $T60$ value as in the SB case. We have depicted the overall structure of the data generation pipeline in Figure \ref{fig:pipeline}. To ensure the generation of realistic parameters, we have analyzed 4495 real $\boldsymbol{T60}$ values from \cite{sridhar2021icassp} in six frequency bands $\boldsymbol{\omega_{n}} = \{125, 250, 500, 1k, 2k, 4k\}$ in Hz. We have modeled each band as an independent $Gamma(\boldsymbol{\alpha}, \boldsymbol{\beta})$ distributions, fitted by minimizing the negative log-likelihood function. We have obtained shape $\boldsymbol{\alpha}=\{ 1.72, 1.62, 1.93, 2.56, 4.17, 2.49\}$ and scale $\boldsymbol{\beta}= \{0.39, 0.24, 0.14, 0.10, 0.09, 0.18\}$. MB RIRs are generated by running one ISM acoustic simulation for each sub-band, passing the sub-band RIRs through a filterbank comprised of a low-pass ($LPF$) for $n=1$, band-pass ($BPF_{\omega_{n}}$) for $n=\{2,...,N-1\}$ and a high-pass filter ($HPF$) for the highest band, and finally summing the filtered RIRs.

\subsection{Receiver directivity RIRs (REC+MB)}
\label{ssec:receiverdirectivity}
In \cite{arakawa2024quantifying}, the augmented reality glasses' directivity and the acoustical shadow of the head were shown to be relevant for speech separation and recognition. Likewise, SE from in-ear or headset devices may also benefit from modeling this directivity. Although there might be individual differences between the device types, they all share the characteristic of being placed near the human ear. We have modeled the receiver directivity in this scenario by applying a Head-Related Transfer Function (HRTF) to every reflection. To this end, we have used the same methodology as in \cite{guso2023objective}: to obtain the Spherical Harmonics (SH) expansion of the RIRs from MASP and then to use a Bilateral Magnitude Least Squares (BiMagLS) \cite{engel2021improving} decoder that implicitly applies the HRTFs, here focusing our evaluation to the left ear and using a set of normal hearing HRTFs of a Neumann KU100 dummy head.

\subsection{Source directivity RIRs (SRC+REC+MB)}
\label{ssec:sourcedirectivity}
In SE, sources are always speech so we have applied its average directivity by taking the radiation pattern from \cite{leishman2021high} and converting it into the azimuth and elevation lookup table $D_{\phi, \theta}$. As depicted in Figure \ref{fig:pipeline}, $D_{\phi, \theta}$ can be applied in the $SH$ pipeline right after the ISM method and prior to the filterbank rendering and summation. We have weighted the amplitude of each ISM reflection with the closest $D_{\phi,\theta}$. Both angles have been obtained through acoustical reciprocity \cite{samarasinghe2017acoustic}: swapping $\boldsymbol{src}$ and $\boldsymbol{rec}$ coordinates in the ISM we obtain a list of reciprocal reflections whose angle with respect to the receiver can be computed directly. Each reciprocal reflection angle with respect to the receiver is equivalent to the emission angle (with respect to the source) in the original ISM.

\subsection{Mesh-based RIRs (SoundSpaces, SSPA)}
\label{ssec:soundspaces}
The SSPA data set contains only 103 scenes, which is much less than scenes in SB. Nevertheless, we have included SSPA in this evaluation as a well-tested reference from fields like navigation \cite{chen2022soundspaces} and 3D SE \cite{marinoni2023overview}. To our knowledge, it is the only publicly-available dataset which can be compared to REC+MB because it also uses Ambisonics for rendering into binaural and multiband absorption coefficients. It additionally provides complex geometries by path-tracing through 3D meshes. To keep the amount of RIRs consistent, we have taken 60k RIRs, also focusing on the left channel of the binaural downmix as in REC+MB and SRC+REC+MB.

\vspace{2cm}

\section{Experimental Setup}
\label{sec:pagestyle}

Given the speech, noise and the six RIR datasets described above, we have trained six  DeepFilterNet3 \cite{schroter2023deepfilternet3} SE models -- one for every RIR dataset-- using its default hyperparameters except for a smaller maximum batch size of 38 in order to fit our GPUs and a $p\_reverb$ of 1 (to apply a RIR to every utterance). We have chosen DeepFilterNet3 because it has SOTA performance while being open source and having real-time capabilities. After training for 117 to 120 epochs (depending on the early stopping) our evaluation has been three-fold: firstly, we have applied the models to monoaural noisy utterances convolved with real RIRs and computed a set of intrusive and non-intrusive metrics. Secondly, we have used a few processed samples to evaluate subjectively with a MUSHRA listening test \cite{series2014method}. Thirdly, we have applied the models to the $headset$ and $speakerphone$ DNS5 test sets to address the effects of directivity. Since these test sets contain real mixtures (no sepparate clean speech ground truth is available) metrics are restricted to be non-intrusive.

\begin{table*}[!ht]
\setlength{\tabcolsep}{7pt}
\renewcommand{\arraystretch}{1.5} 
    \caption{Evaluation results on real RIRs using DNS5 read speech. Mean values $\pm$ standard deviation. The higher the better ($\Uparrow$) except for Log-Spectral Distance and Mel-Cepstral Distortion ($\Downarrow$). SQUIM \cite{kumar2023torchaudio}, NISQA  \cite{mittag2021nisqa} and DNSMOS \cite{reddy2021dnsmos} are non-intrusive neural-based approximations of the metrics.}
    \centering
    \begin{tabular}{c c c c c c c c}

         && DNS5 & SB & MB & REC+MB & SRC+REC+MB & SSPA \\ 
             \hline \hline
         \multicolumn{8}{c}{Intrusive} \\  

        $\Delta SISDR$ & $\Uparrow$ & $1.105_{\pm 3.50}$ & $1.335_{\pm 3.49}$ & $1.361_{\pm 3.45} $ & $\textbf{1.965}_{\pm 4.52}$ & $1.9_{\pm 4.41}$ & $0.888_{\pm 3.27}$ \\ \hline

        $\Delta SDR$ &$\Uparrow$ & $2.817_{\pm 3.17}$ & $3.298_{\pm 3.02}$ & $\textbf{3.328}_{\pm 3.00}$ & $2.674_{\pm 3.19}$ & $2.706_{\pm 3.17}$ & $2.025_{\pm 3.11}$ \\ \hline
        
        $\Delta LSD$ &$\Downarrow$ & $1.213_{\pm 1.41}$ & $1.34_{\pm 1.42}$ & $1.26_{\pm 1.40}$ & $0.841_{\pm 1.43}$ & $\textbf{0.778}_{\pm 1.43}$ & $1.105_{\pm 1.43}$ \\ \hline
        
        $\Delta MCD$&$\Downarrow$ & $-1.434_{\pm 2.00}$ & $-1.479_{\pm 1.98}$ & $-1.504_{\pm 1.99}$ & $\textbf{-1.542}_{\pm 1.97}$ & $-1.533_{\pm 1.98}$ & $-1.454_{\pm 2.00}$ \\ \hline
        
        $\Delta PESQ $&$\Uparrow $ &  $0.52_{\pm 0.47}$  &  $0.583_{\pm 0.47}$  &  $0.593_{\pm 0.47}$  &  $0.595_{\pm 0.46}$  &  $\textbf{0.609}_{\pm 0.46}$  & $0.51_{\pm 0.45}$ \\ \hline
        
        $\Delta STOI$&$\Uparrow$ & $0.088_{\pm 0.07}$ & $0.101_{\pm 0.07}$ & $\textbf{0.102}_{\pm 0.07}$ & $0.101_{\pm 0.06}$ & $0.101_{\pm 0.06}$ & $0.082_{\pm 0.06}$ \\ \hline
        
        $\Delta PhonSim$&$\Uparrow$ & $0.051_{\pm 0.15}$ & $0.058_{\pm 0.15}$ & $0.058_{\pm 0.15}$ & $\textbf{0.063}_{\pm 0.15}$ & $0.0626_{\pm 0.15}$ & $0.061_{\pm 0.15}$ \\ \hline
        
        $\Delta SpkSim$&$\Uparrow$ & $-0.055_{\pm 0.13}$ & $-0.058_{\pm 0.14}$ & $-0.057_{\pm 0.13}$ & $-0.034_{\pm 0.12}$ & $\textbf{-0.03}_{\pm 0.11}$ & $-0.053_{\pm 0.13}$ \\ \hline
        
        $\Delta BertSim$&$\Uparrow$ & $0.081_{\pm 0.07}$ & $0.086_{\pm 0.07}$ & $0.087_{\pm 0.07}$ & $0.089_{\pm 0.07}$ & $\textbf{0.09}_{\pm 0.07}$ & $0.082_{\pm 0.07}$ \\ \hline \hline
         \multicolumn{8}{c}{Non-intrusive} \\  

         $\Delta SISDR_{squim}$ & $\Uparrow$ & $3.793_{\pm 5.87}$ & $4.456_{\pm 5.71}$ & $\textbf{4.544}_{\pm 5.67}$ & $3.814_{\pm 5.65}$ & $4.038_{\pm 5.63}$ & $2.291_{\pm 5.61}$ \\ \hline

       $\Delta MOS_{squim}$&$\Uparrow$ & $\textbf{0.547}_{\pm 0.71}$ & $0.537_{\pm 0.71}$ & $0.541_{\pm 0.71}$ & $0.523_{\pm 0.70}$ & $0.506_{\pm 0.71}$ & $0.541_{\pm 0.70}$ \\ \hline
        
         $\Delta MOS_{dnsmos}$&$\Uparrow$ & $0.884_{\pm 0.53}$ & $0.935_{\pm 0.54}$ & $\textbf{0.937}_{\pm 0.53}$ & $0.895_{\pm 0.53}$ & $0.906_{\pm 0.53}$ & $0.811_{\pm 0.52}$ \\ \hline
              
       $\Delta MOS_{nisqa}$&$\Uparrow$& $1.158_{\pm 0.73}$ & $1.198_{\pm 0.73}$ & $\textbf{1.224}_{\pm 0.72}$ & $1.109_{\pm 0.72}$ & $1.099_{\pm 0.72}$ & $0.977_{\pm 0.73}$ \\ \hline

         $\Delta PESQ_{squim}$&$\Uparrow$ & $0.636_{\pm 0.62}$ & $0.667_{\pm 0.59}$ & $\textbf{0.678}_{\pm 0.59}$ & $0.613_{\pm 0.58}$ & $0.637_{\pm 0.58}$ & $0.525_{\pm 0.57}$ \\ \hline

        $\Delta STOI_{squim}$&$\Uparrow$ & $0.06_{\pm 0.08}$ & $0.07_{\pm 0.08}$ & $\textbf{0.071}_{\pm 0.08}$ & $0.063_{\pm 0.08}$ & $0.066_{\pm 0.08}$ & $0.048_{\pm 0.08}$ \\ \hline \hline
        
             \multicolumn{8}{c}{Subjective} \\  
           $ MUSHRA$&$\Uparrow$ & $33.59_{\pm 20.3}$ & $39.15_{\pm 22.6}$ & $48.05_{\pm 22.8}$ & $42.65_{\pm 22.5}$ & $\textbf{48.39}_{\pm 23.0}$ & $32.49_{\pm 21.7}$ \\ \hline \hline

    \end{tabular}
    \label{tab:res_real}
\end{table*}

\begin{table}[!hb]
\renewcommand{\arraystretch}{1.5}
\caption{Non-intrusive evaluation results on real noisy and reverberant recordings from the DNS5 \textit{headset} and \textit{speakerphone} sets. \textit{HDS} stands for \textit{headset} and \textit{SPK} for \textit{speakerphone}. The higher the better for all metrics.}
\centering
\scalebox{0.77}{
\begin{tabular}{c c c c c c c c}
      & &DNS5 & SB & MB & REC+MB & SRC+REC+MB & SSPA  \\ 
         \hline 

     ${SISDR_{squim}}$ & \begin{tabular}{@{}c@{}}\textit{HDS}\\\textit{SPK}\end{tabular} & \begin{tabular}{@{}c@{}}15.65\\16.81\end{tabular} & \begin{tabular}{@{}c@{}}15.75\\16.86\end{tabular} & \begin{tabular}{@{}c@{}}\textbf{15.80}\\ \textbf{17.05}\end{tabular} & \begin{tabular}{@{}c@{}}15.35\\16.65\end{tabular} & \begin{tabular}{@{}c@{}}15.42\\16.82\end{tabular} & \begin{tabular}{@{}c@{}}14.21\\15.51\end{tabular} \\ \hline

     $MOS_{squim}$ & \begin{tabular}{@{}c@{}}\textit{HDS}\\\textit{SPK}\end{tabular} & \begin{tabular}{@{}c@{}}3.92\\4.15\end{tabular} & \begin{tabular}{@{}c@{}}3.89\\4.15\end{tabular} & \begin{tabular}{@{}c@{}}3.90\\\textbf{4.15}\end{tabular} & \begin{tabular}{@{}c@{}}3.91\\4.14\end{tabular} & \begin{tabular}{@{}c@{}}\textbf{3.94}\\4.14\end{tabular} & \begin{tabular}{@{}c@{}}3.91\\4.14\end{tabular} \\ \hline

     $MOS_{dnsmos}$ & \begin{tabular}{@{}c@{}}\textit{HDS}\\\textit{SPK}\end{tabular}& \begin{tabular}{@{}c@{}}3.01\\3.03\end{tabular} & \begin{tabular}{@{}c@{}}3.04\\3.05\end{tabular} & \begin{tabular}{@{}c@{}}\textbf{3.05}\\\textbf{3.06}\end{tabular} & \begin{tabular}{@{}c@{}}3.04\\3.03\end{tabular} & \begin{tabular}{@{}c@{}}3.02\\3.04\end{tabular} & \begin{tabular}{@{}c@{}}3.01\\2.99\end{tabular} \\ \hline

     $MOS_{nisqa}$ & \begin{tabular}{@{}c@{}}\textit{HDS}\\\textit{SPK}\end{tabular} & \begin{tabular}{@{}c@{}}3.54\\3.77\end{tabular} & \begin{tabular}{@{}c@{}}3.64\\3.82\end{tabular} & \begin{tabular}{@{}c@{}}\textbf{3.68}\\\textbf{3.82}\end{tabular} & \begin{tabular}{@{}c@{}}3.61\\3.77\end{tabular} & \begin{tabular}{@{}c@{}}3.56\\3.76\end{tabular} & \begin{tabular}{@{}c@{}}3.43\\3.62\end{tabular} \\ \hline

     $PESQ_{squim}$ & \begin{tabular}{@{}c@{}}\textit{HDS}\\\textit{SPK}\end{tabular}& \begin{tabular}{@{}c@{}}2.58\\2.64\end{tabular} & \begin{tabular}{@{}c@{}}2.59\\2.63\end{tabular} & \begin{tabular}{@{}c@{}}\textbf{2.62}\\\textbf{2.64} \end{tabular} & \begin{tabular}{@{}c@{}}2.55\\2.56\end{tabular} & \begin{tabular}{@{}c@{}}2.54\\2.59\end{tabular} & \begin{tabular}{@{}c@{}}2.45\\2.49\end{tabular} \\ \hline

     $STOI_{squim}$ &\begin{tabular}{@{}c@{}}\textit{HDS}\\\textit{SPK}\end{tabular} & \begin{tabular}{@{}c@{}}0.92\\0.94\end{tabular} & \begin{tabular}{@{}c@{}}0.92\\0.94\end{tabular} & \begin{tabular}{@{}c@{}}0.92\\0.94\end{tabular} & \begin{tabular}{@{}c@{}}0.92\\0.94\end{tabular} & \begin{tabular}{@{}c@{}}0.92\\0.94\end{tabular} & \begin{tabular}{@{}c@{}}0.91\\0.93\end{tabular} \\ 

\end{tabular}
}
\label{tab:res_spk}
\end{table}

\subsection{Objective evaluation}
\label{ssec:objective}

Speech and noise for these simulations has been taken from the DNS5 test set splits. More precisely, we have taken 10k high quality read speech samples from the VCTK \cite{veaux2017cstr} and LibriVox \cite{panayotov2015librispeech} subsets (7k and 3k respectively). We have used 397 real RIRs from the ACE \cite{eaton2015ace}, MIT IR Survey \cite{traer2016statistics}, Openair \cite{murphy2010openair}, BUT Reverb \cite{szoke2019building} datasets and the RIRs from SLR28 \cite{ko2017study} that were not used during training. Noises and RIRs have been reused as many times as necessary to fit the speech test set size.

From each speech sample we have taken a non-silent four-seconds chunk
$y$, convolved it with a real RIR $h$ and summed it to a noise sample $n$ using $SNR=\mathcal{U}(0, 30)$, obtaining noisy and revereberant  $x$. We have made sure that time synchronicity between $x$ and $y$ is kept, but when addressing this we have found that certain real RIR onsets are harder to detect than their non-noisy synthetic counterparts, so the vicinity of the highest amplitude has been checked for any earlier peak that surpassed a -6dB threshold. We have also found that the broadly used correlation-based synchronization method is less robust to RIR noise than our thresholding heuristic. 

Note that in the traditional signal model $x=(y*h)+n$, noise can be non-reverberant while the speech is, which can constitute a limitation despite being broadly used in most of SE literature. To further investigate this, we have also conducted the evaluation using $x=(y+n)*h$, a signal model in which both speech and noise have a more spectrally-coherent reverberation at the expense of using the exact same RIR (and therefore placing speech and noise at the exact same position in the simulation space, which is deemed unfeasible). Results on this alternative signal model are not reported here for the sake of brevity, but were found to be very similar to the results from the more traditional $x=(y*h)+n$ we report below.

Regarding metrics, we have used almost all of the ones from the URGENT Challenge \cite{zhang2024urgent} with the addition of Scale Invariant Signal-to-Distortion Ration ($SISDR$) and $SISDR_{squim}$, $PESQ_{squim}$, $MOS_{squim}$ and $STOI_{squim}$ from \cite{kumar2023torchaudio}. Because reverberation can mislead some metrics' values and our dataset is the first dataset of this kind that we are aware of, we are reporting the results as increments with respect to the noisy and revereberant signal. Given DeepFilterNet3 model $f$ and clean speech estimate $\hat{y}=f(x)$, we report evaluation metrics $m$ increments as $\Delta m=m(\hat{y})-m(x)$ for non-intrusive metrics and $\Delta m=(m(\hat{y}, y)-m(x, y))$ for the intrusive ones that require the reference $y$. For DeepFilternet3 absolute metrics on the commonly-used Voicebank+Demand test set we refer the reader to \cite{schroter2023deepfilternet3}.

\subsection{Subjective evaluation}
\label{ssec:subjective}
In order to further validate the objective evaluation, we have picked eight reverberant synthetic speech and noise mixtures that use real RIRs from the same set as in the objective evaluation and have conducted an online MUSHRA listening test using \cite{barry2021go, series2014method}, taking $y$ as 48kHz high-quality clean reference and hidden reference, $x$ as anchor and the six different model estimates $\hat{y}$ using a scale from 0 or Bad to 100 or Excellent. All 33 participants have declared to be expert listeners (professional audio producers or researchers with experience doing listening tests). Only one subject had to be excluded during the post-screening due to misjudging the anchor.

\subsection{Computational requirements}
Due to the sequential nature of the ISM and the computational demands of high-order SH, generating all the RIR datasets simultaneously has required a server with 64-cores of CPU and 250GB of RAM for 10 days. Training all the six DeepFilterNet3 models took 30 days on two A100s. 

\section{Results and discussion}
\label{sec:discussion}
\begin{figure}[t]
    \centering
    \centerline{\includegraphics[trim=25 10 10 10, width=\columnwidth, clip]{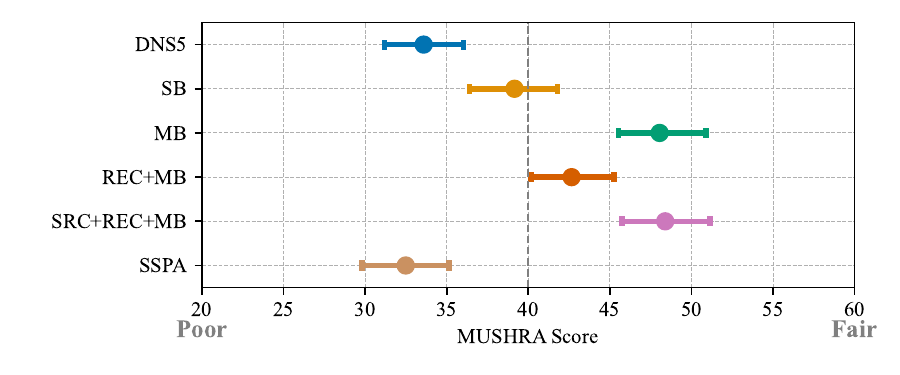}}
    \caption{MUSHRA scores mean and 95\% confidence intervals for each model.}
    \label{fig:sub_results} 
\end{figure}

Subjective results of the enhanced speech and noise mixtures convolved with real RIRs are depicted on Figure \ref{fig:sub_results}.

To compare the conditions, we used a pairwise t-test with a significance threshold of $p<0.05$. MUSHRA scores show no statistically significant difference between DNS5 and SSPA (t-test: $p=0.55$) which points out that variety and quantity of rooms are factors to take into account on top of RIR complexity --note that SSPA additionally models complex room geometries and furniture but despite having the same number of RIRs than SB, it contains much fewer different rooms. SB significantly outperforms the DNS5 baseline (t-test: $p=0.003$), which highlights  the  importance of extending the sampling rate from 16kHz to 48kHz. 

All models apart from SSPA have received significantly better scores than DNS5 baseline (SB: $p=3\cdot10^{-3}$, MB: $p=7\cdot10^{-14}$, REC+MB $p=8\cdot10^{-7}$ and SRC+REC+MB $p=2\cdot10^{-14}$), indicating that both the higher sampling rate  and adding MB, SRC and REC features into the acoustic simulation improves the speech enhancement. Interestingly, there are no significant differences between MB and SRC+REC+MB (t-test: $p=0.86$) which suggests that modeling directivity does not degrade monaural SE performance when evaluated on real RIRs. We can't assess SRC and REC directivities with the real RIR test set, but note that MB and SRC+REC+MB outperform the rest and that all multiband RIR datasets perform similarly. The scores are presented in Table \ref{tab:res_real}, in addition to objective intrusive and non-intrusive results.

Similar to the MUSHRA scores, the objective results on the same set (the real RIRs test set) indicate that MB RIRs outperform the rest. However, there is no consensus among all intrusive metrics, nor between intrusive and non-intrusive metrics. While most intrusive metrics follow the subjective scores and show a higher performance of MB and SRC+REC+MB datasets, $SISDR$, $MCD$ and Phonetic Similarity ($PhonSim$) metrics suggest to use REC+MB. Non-intrusive metrics agreement is higher than in the intrusive ones, with MB slightly outperforming the rest (t-test: $p<0.05$), perhaps due to a bias in NISQA and SQUIM training data but interestingly, towards MB instead of SB.


Comprehensive results for the DNS5 $headset$ and $speakerphone$ are shown in Table \ref{tab:res_spk}. A priori, one would expect SRC+REC+MB to outperform the rest for $headset$, but this is not the case. For both $headset$ and $speakerphone$,  most non-intrusive metrics show MB models slightly outperforming the rest as they did for the real RIR test set, but differences are not statistically significant (e.g. t-test: $p=0.9$ for  $SISDR_{squim}$ between MB and SB). Therefore, we can not find evidence of benefits from directivity modeling.

Although in \cite{arakawa2024quantifying} they showed that directivity could be exploited even in monoaural models, we could not replicate the results, either because of the limitations of NISQA and SQUIM metrics, or due to limitations in the DNS5 test set. In both cases, the lack of a clear improvement when applying directivity brings into question if MB RIRs perform better thanks to being more domain consistent or just because they are more diverse \cite{jeon2024does} than SB. We would like to address this in future work, perhaps by comparing acoustically implausible random $\boldsymbol{T60}$ RIRs with the ones from Section \ref{ssec:multiband}.

Overall, we recommend to train SE models with high sampling rate multiband RIRs, potentially incorporating SRC and REC directivites depending on the use case. We make both MB and SRC+REC+MB available at \href{https://doi.org/10.5281/zenodo.15773093}{\textit{https://doi.org/10.5281/zenodo.15773093}}.

\vspace{0.7cm}

\section{Conclusions}
\label{sec:conclusions}

We have presented an evaluation of three Room Impulse Response (RIRs) generation techniques using the state of the art real-time capable Speech Enhancement (SE) model DeepFilterNet3. Specifically, we have shown that the idea of extending the RIR coverage to frequency dependent acoustic absorption coefficients --which has been shown to be successful for keyword spotting-- can also benefit modern SE models. We have also found that the amount and variability of RIRs can be as important as the model complexity, showing that a lot of ISM-based rooms can outperform fewer mesh-based rooms (SoundSpaces) when evaluated on real RIRs. Besides, we have shown that source and receiver directivities do not degrade performance in the monoaural SE task and also have the potential of increasing SE performance but could not provide strong evidence of the improvement. In conclusion, we recommend training with a varied 48kHz multiband RIRs dataset: MB-RIRs.



\clearpage
\bibliographystyle{IEEEtran}
\bibliography{refs25}

\end{document}